\documentstyle[prb,aps,multicol,epsfig]{revtex}
\begin{document}
\draft
\newcommand{\be}{\begin{equation}}
\newcommand{\ee}{\end{equation}}
\newcommand{\ba}{\begin{eqnarray}}
\newcommand{\ea}{\end{eqnarray}}


\def\inseps#1#2{\def\epsfsize##1##2{#2##1} \centerline{\epsfbox{#1}}}

\def\top#1{\vskip #1\begin{picture}(290,80)(80,500)\thinlines \put(65,500){\line( 1, 0){255}}\put(320,500){\line( 0, 1){
5}}\end{picture}}
\def\bottom#1{\vskip #1\begin{picture}(290,80)(80,500)\thinlines \put(330,500){\line( 1, 0){255}}\put(330,500){\line( 0, -1){
5}}\end{picture}}

\title{Bosonization Theory of Excitons in One-Dimensional
Narrow Gap Semiconductors}
\author{H. C. Lee $^{1}$ and   S. -R. Eric Yang$^{2}$ }
\address{BK21 Physics Research Division and Institute of Basic Science, 
Sung Kyun Kwan University, Suwon, 440-746
Korea$^{1}$
\\ and \\
Department of Physics, Korea University, Seoul, Korea$^{2}$}
\maketitle
\draft
\begin{abstract}
Excitons in one-dimensional narrow
gap semiconductors of anti-crossing
quantum Hall edge states are investigated using a bosonization method.
The excitonic states are studied by mapping the problem  
into a non-integrable sine-Gordon type model.
We also find that many-body interactions
lead to a strong enhancement of the band gap. 
We have estimated when an  
exciton instability may occur.

\end{abstract}
\thispagestyle{empty}
\pacs{{\rm PACS numbers}: \hspace{0.05in} 71.27.+a, 71.35.-y, 71.35.Cc}
\begin{multicols}{2}
One-dimensional (1D) narrow gap semiconductors can be realized in  anti-crossing
quantum Hall edge states\cite{kang} and nanotubes\cite{nano}.  
The gaps in these systems are {\it single} particle gaps and {\it not} many body
gaps.  Theoretically, they provide the  unusual condition that the bare band gap $t$ is 
much {\it smaller} than the characteristic Coulomb energy scale $E_c$.  Moreover,
strong quantum fluctuations are present in these
systems, reflecting the 1D character.  The 
ground state of these systems may be unstable against the
spontaneous formation of excitons if the exciton binding energy exceeds the 
band gap\cite{halperinrice,sham}.  In three dimensional semiconductors  excitons 
can be treated  successfully by solving 
the Bethe-Salpeter equation \cite{sham}.
However, in the strong coupling regime of 1D systems,  
the perturbative approaches
are not expected to be reliable due to large quantum
fluctuations.  
If  the Coulomb
scale is much larger than the gap one might naively expect that  exciton
instability would occur.
However, this simple picture neglects screening which is expected to be large
due to the smallness  of the gap.
It is unclear whether a bound state of an "electron" and a 
"hole"
can exists in the presence of strong quantum fluctuations.
Not much is known about the physics of excitons in 
1D narrow gap semiconductors. 

This problem can be addressed within
a bosonization approach, which is applicable as long as the the characteristic
energy scale of the problem is  smaller than the band width $W$.
Thus bosonization provides  a natural framework for studying
1D excitons.
In this work we will consider anti-crossing
of quantum Hall edge states in the barrier region of between 
two 2D electron gases.  Assume that the applied magnetic field is sufficiently 
strong that the system is spin-polarized with the 
filling factor  $\nu=1$ and that the Fermi level is in the gap of anti-crossing 
edge states.  
We find that  the problem can be mapped into a sine-Gordon (sG)
model with an extra term
representing the long range Coulomb interaction.  Due to this extra
term the model becomes {\it non-integrable} and an exact solution is unavailable.
However, the non-linear cosine term of this sG model can be expanded provided that 
$\delta_{ex}\sim \frac{v/v_0}{ \ln{W/t}}<1$, where 
$v$ is the Fermi velocity, $v_0=e^2/\epsilon \hbar$, $W=\hbar v/a$, and $a$
is the smallest length scale in the problem\cite{com}.
In this regime 
a perturbative theory may be applied to calculate excitonic energies.
Our work also indicates that a large enhancement of band gap occurs.
Our perturbative approach suggests an approximate estimate for when an  
exciton
instability may occur: $\delta_{ex} > 1$ .
(We cannot
exclude the possibility that the  higher order
corrections neglected in our perturbative calculation may 
work against such an instability). 

The system is modeled by  the following Hamiltonian\cite{wen}
\ba
\label{H1}
H&=&H_0+H_{{\rm coul}}+H_{t},\nonumber \\
H_0&=&v \int dx \Big[-i  \psi^\dag_{R }\partial_x \psi_{R }+
i \psi^\dag_{L } \partial_x \psi_{L } \Big] \nonumber \\
&=&
 \pi v \int dx  \Big[ \rho_{R }^2+ \rho_{L }^2 
\Big] \\
H_{{\rm coul}}&=& \int dx dy \Big[ \frac{V(x-y)}{2} \big(
 \rho_R(x) \rho_R(y)+\rho_L(x) \rho_L(y) \big) \nonumber \\
 &+& V(x-y) \big(
\rho_L(x) \rho_R(y)\big) \Big]\nonumber \\
H_{t}&=&-t \int dx \Big[ 
\psi_{R }^\dag(x) \psi_{L }(x)+ {\rm H.C} \Big],
\ea
The operators $\psi_{R } (\psi_{L })$ is the right-moving (left-moving) 
edge {\em electron} operators.
$\rho_{R }=:\psi^\dag_{R } \psi_{R }:$ is the (normal-ordered) right-moving
edge electron density operator ($\rho_{L }$ is similarly defined).
In practice, we can take $V(x)=\frac{e^2}{\epsilon}\frac{1}{
\sqrt{x^2+a^2}}$ and 
the Coulomb matrix element is $V(k)=\frac{ 2 e^2}{\epsilon} K_0( a |k|)$.
The tunneling between the right-moving and left-moving  is modeled by
$H_{t}$.  The tunneling amplitude $t$ is assumed to be significant only
near the Fermi wave vector $k_F$(The Fermi wavevector is set to zero from now on). The   
cut-off value  of the wave vector  
is  $1/a$.
Note that a single particle gap  opens up near the
Fermi points due to this tunneling term. 
The commutation relations of density operators in momentum space are given by
\cite{wen} $
[\rho_{R }(q),\rho_{R }(q^\prime)]=-\frac{q}{2\pi} 
\delta_{q+q^\prime},
[\rho_{L }(q),\rho_{L }(q^\prime)]=+\frac{q}{2\pi} 
\delta_{q+q^\prime}$
The interacting electron systems can be bosonized in a standard way. 
The explicit relations between the  electron  operators and
the bosonic variable $\phi_{R }$, $\phi_{L }$ are  $
\rho_{R }=\frac{1}{2\pi} \partial_x \phi_{R },
\rho_{L }=\frac{1}{2\pi} \partial_x \phi_{L },
\psi_{R }=\frac{1}{\sqrt{2\pi a}} e^{ i \phi_{R }},
\psi_{L }=\frac{1}{\sqrt{2\pi a}} e^{ -i \phi_{L }}$.
It is convenient to introduce 
the conjugate fields $\phi_{\pm }$:
$\phi_{\pm }\equiv \phi_{R } \pm \phi_{L }$.
The effective bosonized action in imaginary time reads
\ba
\label{action2}
S&=&\int dx d\tau \Big[ \frac{v}{8\pi}\,\big( (\partial_x \phi_{+})^2+
(\partial_x \phi_{-})^2 \big) + \frac{i}{4\pi } \partial_\tau \phi_+ 
\partial_x \phi_-\Big] \nonumber \\
&+& \frac{1}{8\pi^2} \int dx dy d\tau V(x-y) \partial_x \phi_+(x)  \partial_y
\phi_+(y) \nonumber \\
&-&\frac{t}{2\pi a} \int dx d\tau
 \Big[ e^{i\phi_+(x)}+{\rm H.C}\Big].
\ea
The $\phi_-$ can be integrated out, leaving us with
\ba
\label{action}
S&=&\frac{1}{8\pi }\int \frac{dk d \omega}{(2\pi)^2}
 \Big[v k^2 (1+ \frac{V(k)}{\pi v})
 +\frac{\omega^2}{v}  \Big] \nonumber \\
 &\times& \,\phi_+(\omega,k) \phi_+(-\omega,-k)
\nonumber \\
&-& \frac{t}{\pi a}\, \int dx d\tau \cos(\phi_+(x,\tau)).
\ea

The above action looks very similar to the sG model\cite{rajaraman},
except for
the momentum-dependent Coulomb interaction $V(k)$. If $V(k)$ were momentum
independent (local interaction in real space), the action would be exactly 
that of sG model. The Euclidean action of the sine-Gordon model is
given by\cite{rajaraman,lukyanov,zam2}
\be
\label{sg}
A_{SG}=\int d^2x \Big[ \frac{1}{16\pi} (\partial_\mu \phi)^2-2 \mu \cos(
\beta \phi) \Big],
\ee
where the speed of "light" has been set to unity\cite{light}.
It is useful to show the equivalence between the above sG action
and our action (\ref{action}) 
for the short range interaction ($V(k)=V$ is constant).
This is achieved by the relations
\ba
\phi&=&\phi_+\,[4 (1+ \frac{V}{\pi v})]^{1/4},\;\;
\beta=[4 (1+ \frac{V}{\pi v})]^{-1/4},\nonumber \\
\mu&=&\frac{t}{\pi a  v (1+V/\pi v)^{1/2}}.
\ea
From the above relations it is clear that the {\it strong} coupling regime 
of the original electron system (large $V$) is mapped to the {\it weak} coupling
regime (small $\beta$)of sG model.
 
Before we investigate the long-range case it is instructive to review
some known the physical properties of sG model, which is exactly solvable.
The spectrum of sG model consist of the breathers ($B_n$), the soliton, and 
the anti-solition. The breathers are the bound states of the soliton and 
anti-soliton \cite{rajaraman,lukyanov,zam2}.
The lightest  bound state $B_1$ coincides with the fundamental boson $\phi$ of
the sG model (\ref{sg}), and it is
associated with the perturbative treatment.
The number of breather is given by
$n=1,2,3,\ldots< 1/\xi$,
where  $\xi=\frac{\beta^2}{1-\beta^2}$.  
Note that even for infinitesimally small short 
range repulsion $V$, at least one breather exists. The breather energy (excitation energy)
is given by 
\be
\label{breathermass}
m=2M \sin \frac{\pi \xi}{2},
\ee
where $M$ is the energy of solition.  The parameter  $\mu$ and the soltion energy
$M$ are related through
\be
\label{mass}
\mu=\frac{\Gamma(\beta^2)}{\pi \Gamma(1-\beta^2)}\left
[ M \frac{\sqrt{\pi} \Gamma(\frac{1+\xi}{2})}{
2 \Gamma(\frac{\xi}{2})}\right]^{2-2 \beta^2}.
\ee
Note that in the $\beta \to 0 $ limit,
$M \sim v_\rho^2\mu/\beta^2 \to \infty$ and 
$ m \sim v_\rho^2(\mu/\beta^2) \beta^4 \sim v_\rho^2 \mu \beta^2 \sim t/va $, where "the speed of light" 
$v_\rho$ has been reinstated for
clarity \cite{light}.
In other words, in the large $V$ limit 
the soliton becomes very massive and leads to a 
large enhancement of the band gap.
In contrast,  the lightest breather
energy approaches a constant value given by the coefficient of the cosine term of Eq.(\ref{action}).

In the {\em absence}  of the Coulomb interaction the exact value of the single 
particle gap should be $t$.
However,  the
bosonized action (\ref{action})  contain the factor $a$ , and it is unclear
how this factor disappears in the final result for the gap.
Let us try to understand how this happens.
The crucial fact is that the cosine term of the sG
action (\ref{sg}) is normal ordered and it gives 
the dimension to the cosine operator. 
The exact results Eq.(\ref{breathermass},
\ref{mass})\cite{lukyanov,zam2} were obtained with  
 \be
 \label{2}
\mu^2 \langle :\cos(\beta \phi(x)): :\cos(\beta \phi(y)): \rangle 
\to \frac{1}{2} \frac{\mu^2}{|x-y|^{4 \beta^2}},
\ee
as $ |x-y| \to 0 $.
In the absence of Coulomb interaction, corresponding to $\beta^2=1/2$ ,  the 
same correlation functions (two-point function of tunneling term)
can be computed exactly when expressed in terms of electron
operators
\be
\label{1}
\langle [t \psi^\dag_R(x) \psi_L(x)] [t\psi^\dag_L(y) \psi_R(y)] \rangle
=\frac{t^2}{|x-y|^2}.
\ee
The formal application of bosonization formula to this  electron tunneling
term gives $\frac{t}{\pi a} \cos(\beta \phi)$ as in our action (\ref{action}). 
As it stands  it is not normal ordered, and consequently 
the short-distance singularity must be regularized in the calculation of 
correlation function.
\ba
& &\left( \frac{t}{2\pi a} \right)^2 
\langle e^{i\beta \phi(x)} e^{-i \beta \phi(y)} \rangle= \nonumber \\
& &\left( \frac{t}{2\pi a} \right)^2 \exp\Big[-\beta^2(
\langle \phi(0) \phi(0) \rangle-\langle \phi(x) \phi(y) \rangle )\Big].
\ea
We have to regularize $\langle \phi(0) \phi(0) \rangle$ as
$\langle \phi(0) \phi(a) \rangle=2 \ln \frac{L^2}{a^2}$, 
where $L$ is the system 
size. $L$ always appears  in the definition of propagator but they
are canceled for the physical correlation functions (see below).
Since $\langle \phi(x) \phi(y) \rangle=2 \ln \frac{L^2}{|x-y|^2}$
we obtain
\be
\label{3}
\left( \frac{t}{2\pi a} \right)^2 
\langle e^{i\beta \phi(x)} e^{-i \beta \phi(y)} \rangle=\left( \frac{t}{2\pi a} \right)^2 
\frac{1}{(|x-y|/a)^{4 \beta^2}}.
\ee
Only at $\beta^2=1/2$, the length scale $a$ drops out.  
Comparing  Eq.(\ref{2}) with Eq.(\ref{3}),
the $t$ can identified with $\mu$. 
Then by applying formula  Eq.(\ref{mass}) at $\beta^2=1/2$
, we find that $t = M $, which is the expected result.
The above argument checks the consistency of the bosonization formulation
of our problem.
We emphasize again, the above results holds
only if the Coulomb interaction is absent.

When the long range 
Coulomb interaction is present 
the model becomes {\it non-integrable} and an exact solution is unavailable.
But, under certain conditions, the cosine
term of the action (\ref{action}) can be expanded and excitons may be
studied perturbatively. 
(See the perturbative calculation below).
Expanding the cosine term of (\ref{action}) up to the 4th order, we get an effective action of exciton states.
\ba
\label{approxaction2}
S&\approx & \frac{1}{8\pi}\,\int \frac{dk d\omega}{(2\pi)^2}\,
\Big[ v k^2(1+ \frac{V(k)}{\pi v})+\frac{\omega^2 }{v}+\frac{4 
t }{\hbar a} \Big] \nonumber \\
&\times& \phi_+(\omega,k) \phi_+(-\omega,-k) \nonumber \\
&-&\frac{1}{4!}\frac{t}{\pi \hbar a}\,\int dx d\tau :\phi_+^4:.
\ea
At this point, we note that the parameter $\frac{t}{\pi a}$ 
in action (\ref{action}) should be understood
as a renormalized quantity. The application of bosonization formula to the 
fermion bilinear mass term assumes the implicit normal ordering \cite{mandelstam}.
Such a normal ordering sums the tadpole diagrams\cite{coleman} (see Fig.1) 
and the effect of 
the normal ordering is contained in the cut-off parameter $\frac{1}{2\pi a}$.
Therefore, in the perturbative expansion of cosine term we have to exclude all the 
tadpole type diagrams, and then all other terms of perturbative expansion are
finite and well-defined. 
In analogy with the sG model,
we will 
associate the solition-like
(quantum) solution of (\ref{action}) with the renormalized electrons, and
the breather-like $\phi_+$ mode with the excitonic  electron-hole bound state.

If the quantum correction due to $\phi_+^4$ is neglected, the zeroth order
dispersion
relation of exciton state can be read off  from the  quadratic
part of (\ref{approxaction2}).
\be
E^{(0)}(k)=\hbar \Big[ v^2 k^2 (1+ \frac{v_0}{2v} \ln 
\frac{1}{a |k|})+\frac{4 t v}{\hbar a} \Big]^{1/2}.
\ee
The energy gap is given by 
\be
\label{excitongap}
\Delta^{(0)}_{ex}=\sqrt{\frac{ 4 \hbar t v}{ a}}=2\sqrt{ t W}.
\ee
Next, we compute the correction to the exciton excitation energy due to the
quantum fluctuation of $\phi^4_+$ term. The first such correction appears in 
the second order perturbation of $\phi_+^4$.
Because the $\phi_+^4$term is normal ordered there are no tadpole 
diagrams. Only  the "sunset" diagram (see Fig.1) contributes, 
which is both ultraviolet and
infrared convergent. 
\begin{figure}[b]
\inseps{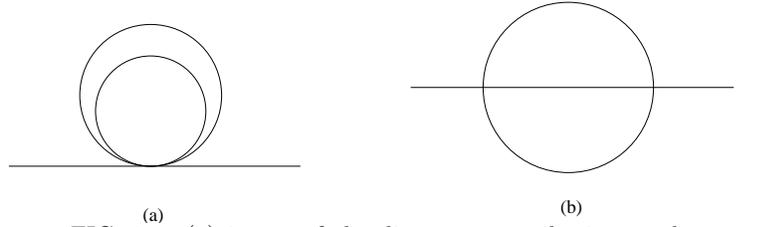}{0.66}
\caption{\label{fig1}
(a) is one of the diagrams contributing to the normal ordering, and 
(b) is the "sunset" diagram. The solid lines are the boson 
propagator $\langle \phi_+ \phi_+ \rangle$.}
\end{figure}
\noindent
It contributes a {\it negative} correction 
to the gap as can be checked by a simple calculation
$e^{-(\Delta_{ex}^{(0)})^2 \phi^2}(1+\lambda^2 \phi^2 \langle \phi^3 \phi^3 \rangle) \sim
e^{-((\Delta_{ex}^{(0)})^2-\lambda^2 \langle \phi^3 \phi^3 \rangle) \phi^2},$
where $\lambda \sim t/\pi \hbar a$.
The symmetry factor of the sun-set diagram is
$\frac{1}{4!} \frac{1}{4!} 4 \cdot 4 \cdot 3 \cdot 2=1/6$.
In terms of self-energy the sunset diagram gives
\ba
\Sigma(i\omega, k)&=& \frac{1}{6}
 \left( \frac{t}{\pi \hbar a} \right)^2
\int d x d \tau [D(x,\tau)]^3  e^{ikx+i\omega \tau}, \nonumber \\
D(x,\tau)&=&\int \frac{d \omega dk}{(2\pi)^2} 
\frac{ v e^{ikx+i\omega \tau}}{\omega^2+v^2 k^2(1+ \frac{\alpha}{2}
  \ln \frac{1}{|k| a})
+\frac{4 t v}{\hbar a}}.
\ea
The excitation energy which includes the above quantum correction is given by
\be
\Delta_{ex}=\Big[\frac{ 4 \hbar t v}{ a}- 8 \pi \hbar^2 v 
\Sigma(0,0)\Big]^{1/2}.
\ee
Write the above equation in the form
\be
\label{newgap}
\Delta^2_{ex}=(\Delta^{(0)}_{ex})^2 \,\Big[1-
\frac{ \Sigma(0,0)}{t/(2\pi \hbar a)}\Big].
\ee
The second term in the bracket of Eq.(\ref{newgap})
is the quantum correction in the dimensionless
form.
Explicitly, ($y_1,y_2,\eta_1,\eta_2$ are dimensionless).
\ba
&&\delta_{ex}=\frac{ \Sigma(0,0)}{t/(2\pi \hbar a)}=
\frac{1}{4\pi}\,\int d y_1 d y_2 \big[D(y_1,y_2) \big]^3, \nonumber \\
&&D(y_1,y_2)=\int \frac{d \eta_1 d \eta_2}{(2\pi)^2}\,
\frac{e^{i \eta_1 y_1+ i \eta_2 y_2}}{
4+ \eta_1^2 +\eta_2^2 f(|\eta_2|)},
\ea
where $f(|\eta_2|)=1+\frac{v_0}{2v} \ln [\frac{1}{ |\eta_2|}
\sqrt{\frac{W}{ t}}]$.
We may extract the effective expansion parameter  by rescaling.
It is easy to see  that 
\be
\delta_{ex} \sim \frac{\delta \Sigma(0,0)}{t/(2\pi \hbar a)} 
\sim \frac{v/v_0}{ \ln{W/t}}.
\ee
We note that, in the perturbative regime $\delta_{ex} \rightarrow 0$ 
the excitation energy
approaches the value of the coefficient of the cosine term.
This same feature  also appeared in the exactly solvable sine-Gordon model.
Our perturbative treatment for the excition states is thus valid
when  
$\delta_{ex}<1$.
This result suggests the possibility that when $\delta_{ex} > 1$ is satisfied  
exciton
instability may occur.   However, we cannot
exclude the possibility that the  higher order
corrections neglected in our perturbative calculation may 
prohibit such an instability.

The action Eq.(\ref{action}) supports  topological soliton excitations due 
to the topologically inequivalent vacuum of the cosine potential
\cite{rajaraman}. Thus, we can {\it not} use the perturbative expansion of the cosine
term in the study of the solition excitation.
The original sine-Gordon model can be solved exactly, and 
the energy gaps of soliton and breather are known exactly (see
Eq.(\ref{breathermass}), Eq.(\ref{mass}))
\cite{rajaraman,zam2}.   Our action
(\ref{action}) is not exactly solvable, and here we will give only an
estimate of the gap of soliton excitation.
In estimating the gap of soliton excitation, it is sufficient to consider
the static solition.
In our perturbative regime
we may neglect 
$1$ compared to $\frac{V(k)}{\pi v}$.  Then, in real space, 
the energy functional for static solution reads
\ba
E[\phi_+(x)]&=&\frac{1}{8\pi}\int dx dy \frac{e^2}{\epsilon|x-y|} 
\partial_x \phi_+(x)
\partial_y \phi_+(y) \nonumber \\
&+&\frac{t}{\pi a}\,\int dx (1-\cos(\phi_+(x)),
\ea
(A constant term is added to make the expression positive definite).
We define a dimensionless variable $\bar{x}$, $
\bar{x}=x \left(\frac{t}{e^2 \epsilon a} \right)^{1/2}
=x\left(\frac{t}{E_c} \right)^{1/2}\frac{1}{a}$.
Then, the energy functional becomes
\ba
E[\phi_+(x)]&=&\sqrt{ \frac{e^2 t }{\epsilon a}}\,\Big[
\frac{1}{8\pi}\int d\bar{x} d\bar{y} 
\frac{1}{|\bar{x}-\bar{y}|} \partial_{\bar{x}} \phi_+(\bar{x})
\partial_{\bar{y}} 
\phi_+(\bar{y}) \nonumber \\
&+&\,\frac{1}{\pi}\int d\bar{x} (1-\cos(\phi_+(\bar{x}))\Big].
\ea
The coefficient in front of bracket has the dimension of energy and it gives a 
characteristic gap scale of solitions.
The ground state, which sits at a minimum of the cosine potential,
has  zero classical
energy. The soliton excitation connects two adjacent  classical ground
states, and according the above estimate the gap of solition $E_G$ is 
of the order $\sqrt{ \frac{e^2 t }{\epsilon a}}=\sqrt{E_c t}$.
Since we expect $E_c >>t$  the Coulomb interaction 
enhances the value of band gap 
significantly. 

In summary, we have studied  
excitons in 1D narrow gap semiconductors
of two anticrossing  quantum Hall edges.
According to our  perturbative approach the exciton state may lie in the gap
when 
$\delta_{ex} < 1$.  
Our study indicates
that many body interactions enhances the value of band gap 
significantly.   
Our result suggests that an exciton instability
may occur when $\delta_{ex} \sim  1$.  However, 
it is desirable to calculate how the  higher order
corrections neglected in our perturbative calculation may 
change this condition.  
The actual values of 
$v$,  and $t$ in anticrossing edges states are not known well and
it is difficult to estimate precisely the actual value of $\delta_{ex}$\cite{com2}.
Experimental observation of excitons and the  investigation of exciton instability 
would be most interesting.

We are  grateful to Al. B. Zamolodchikov and M.P.A. Fisher for useful comments.
H. C. L. was  supported by the Korea Science and Engineering
Foundation (KOSEF) through the grant No. 1999-2-11400-005-5, and by the 
Ministry of Education through Brain Korea 21 SNU-SKKU Program.
S.R.E.Y was supported by Korea Research Foundation Grant KRF-2000-
015-DP0125.


\end{multicols}

\begin{references}
\bibitem{com}There are several small lengths in this problem: the transverse 
dimension of the quasi-1D system $w$, the magnetic length $\ell$, 
and the width of the barrier $a$.  They are all of order 100 \AA.  In this paper
they will all be denoted by $a$.
\bibitem{light} The explicit expression of the "speed of light" is 
$v_\rho=v (1+V/\pi v)^{1/2}$. Note its dependence on $V$.
\bibitem{com2} In our estimate it is likely that  $0.1< \delta_{ex}<1$
\end{references}
\end{document}